# ASPECTS OF GOND ASTRONOMY

**M.N. Vahia**
*Tata Institute of Fundamental Research, Homi Bhaba Road,*
*Colaba, Mumbai 400 005, India.*
Email: vahia@tifr.res.in

**and**

**Ganesh Halkare**
*Indrayani Colony, Amravati, 444 607, India.*

Abstact: The Gond community is considered to be one of the most ancient tribes of India with a continuing history of several thousand years. They are also known for their largely isolated history which they have retained through the millennia. Several of their intellectual traditions therefore are a record of parallel aspects of human intellectual growth, and still preserve their original flavour and have not been homogenised by the later traditions of India. In view of this, the Gonds provide a special window to the different currents that constitute contemporary India. In the present study, we summarise their mythology, genetics and script. We then investigate their astronomical traditions and try to understand this community through a survey of 15 Gond villages spread over Maharashtra, Andhra Pradesh and Madhya Pradesh. We show that they have a distinctly different view of the sky from the conventional astronomical ideas encountered elsewhere in India, which is both interesting and informative. We briefly comment on other aspects of their life as culled from our encounters with different members of the Gond community.

Keywords: India, Gonds, indigenous astronomy.

## 1 INTRODUCTION

The Gonds are the largest of the Indian tribes, with a population of between 4 and 5 million spread over northern Andhra Pradesh, eastern Maharashtra, eastern Madhya Pradesh, Jharkhand and western Orissa (Fürer-Haimendorf and Fürer-Haimendorf, 1979). While their precise history cannot be dated to a period earlier than AD 890 (Deogaonkar, 2007: 37), their roots are certainly older.

## 2 THE GONDS

### 2.1 The Origin of the Gonds

Mehta (1984: 105-215) has studied the Gonds from different perspectives, and also their history and mythology in detail. Based on linguistic and other data he considers them to be an ancient community, and one of the oldest tribes in India, with their roots going back to a pre-Dravidian arrival in south India around 2000 BC. He identifies later Brahman influences in their stories. Based on ideas of totem poles and other signs of early religion he makes a very strong case to consider them as one of the earliest inhabitants of central India, with the core in the Kalahadi region of Orissa. Interestingly however, the Gonds consider themselves to be later entrants into God's world through the penance of *Shiva's* son *Karta Subal* (Mehta, 1984: 177). It has also been suggested that they were descendants of *Ravan* (Mehta, 1984: 205). Aatram (1989: 141-143) has suggested a connection between the Gonds and the reference to the Kuyevo tribe in the *Rig Veda*.

The history of the Gonds suggests that they occupied large stretches of land in central India and were its primary rulers from AD 1300 to 1600 (Deogaonkar, 2007: 34-55). However, one of the conspicuous aspects of the Gond lifestyle has been that they did not transform from farmers using the simplest farming techniques to an urban, settled population until very recently. Moreover, they did not evolve into a formal civilisation, living in cities, with elaborate trading practices, and become a large non-agricultural population. This may have been due to a lack of any need to create surpluses, conserve resources and rationalise their population groups (e.g. see Vahia and Yadav, 2011). The reasons for this need to be studied separately.

Sociologically, the Gonds ruled large parts of central India before the rise of the Mughal Empire in Delhi. Several forts and other relics from the Gond Kingdom suggest their dominance over central India during this period. The fact that they built forts and not castles also suggests a lack of desire to move from agricultural roots to urbanisation. Their current lifestyle is also indicative of farming traditions rather than aggressive kingdom-building. The impact of acculturation since their original roots and their subsequent integration into respective state linguistic and religious traditions has resulted in a recent strong desire to revive their original traditions and preserve their group identity.

### 2.2 The Geographical Spread of the Gonds

The Gonds are mainly divided into four tribes: Raj Gonds, Madia (Maria) Gonds, Dhurve Gonds and Khatulwar (Khutwad) Gonds. Deogaonkar (2007: 15-16), quoting Mehta (1984), lists the major areas of the Gonds to be:

1. The Bastar region in Madhya Pradesh on the Godavari Basin
2. The Kalahandi region of Orissa
3. The Chandrapur region of Maharashtra
4. The Adilabad region of Andhra Pradesh
5. The Satpuda and Narmada regions of Madhya Pradesh
6. The Raipur region in Madhya Pradesh, including Sambalpur in Chattisgarh, and the Sagar region in Madhya Pradesh
7. The Ellichpur region in the Amravati District of Maharashtra

Their population size has increased from about 100,000 in the 1860s (Deogaonkar, 2007: 23) to about 3.2 million in the 1941 census (Agrawal, 2006: 35) and to 4.1 million in 1961 (Deogaonkar, 2007: 13). Their population as per the 1991 census was 9.1 million (after Wikipedia). Compared to this, the population of India as a whole rose from about 250 million (of undivided India) in 1870 to 360 million in 1950 and 490 million in 1965 (Maddison, 1989: 129). The population of India in 1991 was 850 million (after Wikipedia). The relatively steep increase in their population (which is rising faster than the general population of India) suggests that the Gonds originally lived in low-density population groups over large tracks of land and had a low life expectancy. However, there has been a change in this trend: integration into the larger Indian population, subsequent lifestyle changes and a significant improvement in their general well-being have resulted in increased longevity of the Gond population.

### 2.3 Genetic and Linguistic Data on the Gonds

Genetically the Gonds are a mix of Dravidian and Austro-Asian populations (Balgir, 2006; Gaikwad et al., 2006; Pingle, 1984, Pingle and Fürer-Haimendorf, 1987; Sahoo and Kashyap, 2005), while some genetic markers are unique to this population. In particular, two genetic markers, loci D3S1358 and FGA, show departure from the Hardy-Weinberg equilibrium in the Gond tribe. These are also markedly different from those of seven neighbouring populations (4 tribes and 3 castes—two middle castes and one Deshasth Brahmin caste) (Dubey et al., 2009) indicating that the Gonds have been able to maintain their genetic isolation, with little intermixing with neighbouring tribes.

Linguistic studies of the Gond language show that Gond tribes comprising the Madia-Gond, a hunter-gatherer population, harbour lower diversity than the Marathi tribal groups, which are culturally and genetically distinct. The Proto-Australoid tribal populations were genetically differentiated from castes of similar morphology, suggesting different evolutionary mechanisms operated within these populations. The populations showed genetic and linguistic similarity, barring a few groups with varied migratory histories. The microsatellite variation showed the interplay of socio-cultural factors (linguistic, geographical contiguity) and micro-evolutionary processes. Gond culture and language therefore can be considered isolated, and the level of contamination or modification by interaction with other tribes seems to be low. This is seen from the fact that while they use local names for the numbers 9 and 10, they continue to maintain their original number-name associations for the numbers 1 to 8. This is also reinforced by the fact that they continue to ignore the current Pole Star (Polaris), and do not seem to have a specific name for it (see further discussion on this point below). This evidence of isolation therefore permits us to study their indigenous beliefs without having to allow for cultural contamination.

### 2.4 The Religion and Customs of the Gonds

In religious terms, there are nine distinct groups of gods whose lineages are followed by all Gonds. Their primary god is *Bada Deo* or *Mahadev* (*Pen*) who is conventionally thought to be *Shiva* of the Hindu traditions. But at an operational level, there are nine groups of gods, and these are referred to by numbers (1 to 7, 12 and 16). However, references to twelve gods (from 1 to 12) named simply as *Undidev Saga*, *Randudev Saga* all the way to *Padvendev Saga* (the 10$^{th}$ God), *Pandunddev Saga* (the 11$^{th}$ God) and *Panderdev Saga* (the 12$^{th}$ God) can also be found, and they all have names. Each Gond is a follower of one of the numbered groups of gods. Members belonging to the lineage of the even-numbered group of gods were originally permitted to marry only those belonging to the odd-numbered group of gods, but this tradition is now changing. In addition, the Gonds have further subdivisions by surname and *gotra* (clan).[1] Conventionally there are believed to be 750 distinct *gotras*, a number that is marked on their flag (see Kangali, 1997: 183-185).

According to the 2007 Gondvana Kiran Calendar the Gonds have 24 major festivals, and these are listed in Table 1. The last column only gives the approximate Gregorian month, since synchronisation of solar and lunar months is only done periodically. Consequently, in a specific year, the New and Full Moon may fall in the previous or the following Gregorian month to the one mentioned here.

Gond customs also vary significantly from classical Hindu customs. Conventionally, Gonds bury their dead with the head of the body facing south in most regions, but to the west in some areas. They consider north to be a direction of ill omen that brings disaster. By contrast, south is considered to be a holy direction. This is the reverse of Hindu convention. A small stone marks the location of a burial. However, traditions of creating hero stones closer to home, and common community worship, are also known. In one community, we were also given reference to other gods, which included *Kali, Kankali*, *Maikali*, *Jango*, *Lingo*, *Jari-Mari*, *Maan-ko*, *Tadoba*, *Vagoba*, *Guru* and *Pahandi-Kupar* (Kangali, 1997). Their primary temples worship snakes and *Mahadeo*, but temples dedicated to weapons and other iron tools, and to memorials of *Rani Durgavati,* also can be found. The primary symbol of worship is a complex fertility symbol (Figure 1). It is interpreted as having a feminine representation at the bottom followed by the male *lingam*, and with Earth and the Sun on top, all interconnected in some representation and shown separately on flags etc.

Table 1: Festival Days of the Gonds.

| No | Festival name in Gondi | Festival name | Lunar calendar date | Approximate Gregorian Month |
|---|---|---|---|---|
| 1 | Say Mutholi | Worship of Panch Pavli | Magha Full Moon | January – February |
| 2 | Sambhu Naraka | Shiv Jagran | 2 days prior to Magh New Moon | January – February |
| 3 | Shivam Gavara | Worship of Shica (Shigma) | Fagun 5th day from New Moon | February – March |
| 4 | Khandera | Worship of Meghnath | Fagun 5th day from New Moon | February – March |
| 5 | Ravan Muri | Worship of Ravan | Fagun 5th day from Full Moon | February – March |
| 6 | Mand Amas | Worship of Mand | Fagun New Moon | February – March |
| 7 | Kuvara Bhimal Puja | Worship of Bhivsan | Chaitra Full Moon | March – April |
| 8 | Mata May Puja | Worship of Mata May | Chaitra 5th day from Full Moon | March – April |
| 9 | Nalenj Puja | Worship of the Moon | Chaitra New Moon | March – April |
| 10 | Naya Khana | Festival of new food | Vaishakh 5th day since New Moon | April – May |
| 11 | Budhadev Puja | Worship of Budhadev | Vaishakh Full Moon | April – May |
| 12 | Sajori Bidari | | Jyeshtha Full Moon | May – June |
| 13 | Hariyommat | Worship of fruits and plants | Jyeshtha New Moon | May – June |
| 14 | Thakur Dev Puja | Time for sowing seeds | Akti | May – June |
| 15 | Khut Puja | Worship of Khut | Ashadh Full moon | June – July |
| 16 | Saag Pen Puja | Worship of Saag Pen | Ashadh New Moon | June – July |
| 17 | Naag Panchami | Worship of the Snake, particularly the King Cobra | Shravan 5th day from New Moon | July – August |
| 18 | Saila Puja | Worship through dance | Shravan Full Moon | July – August |
| 19 | Pola | Worship of Pola | Shravan New Moon | July – August |
| 20 | Naya hana | New Food Festival | Bhado 5th day from New Moon | August – September |
| 21 | Navaratra | 9 day festival of worship of Durga | Ashvin 10th day from Full Moon | September – October |
| 22 | Jango – Lingo Lati Puja | Worship of *Jango* and *Lingo* (the Sun and Moon) | Kartik Purnima | October – November |
| 23 | Nagar Puja | Worship of the village | Kartik Purnima | October – November |
| 24 | Kalimay Puja | Worship of Kali Kankali | Paush New Moon | December – January |

## 2.5 The Gond Script

There is significant confusion about the existence of a Gond script and both Deogaonkar (2007: 123) and Mehta (1984: 173) suggest that there is no Gond script at all. However, we came across examples of Gond writing in several places. We found examples of a calendar written in the Gond language (i.e. in Gondi), with the first sheet (Figure 2) discussing the Gond script.[2] In line with the unique features of scripts of the Subcontinent, it also merges the vowels and consonants to create complex signs which require careful reading but can retain subtle aspects of pronunciation.

The page reproduced here in Figure 2, and the calendar, have the writing of names and numbers relating to the calendar in the original script, its transliteration and translation. As an example, we list the days of the week transliterated from Gondi in Table 2, and in Table 3 we list the names of the months. This is claimed to be the original text, but it is not clear when and how the present structure was finalised.

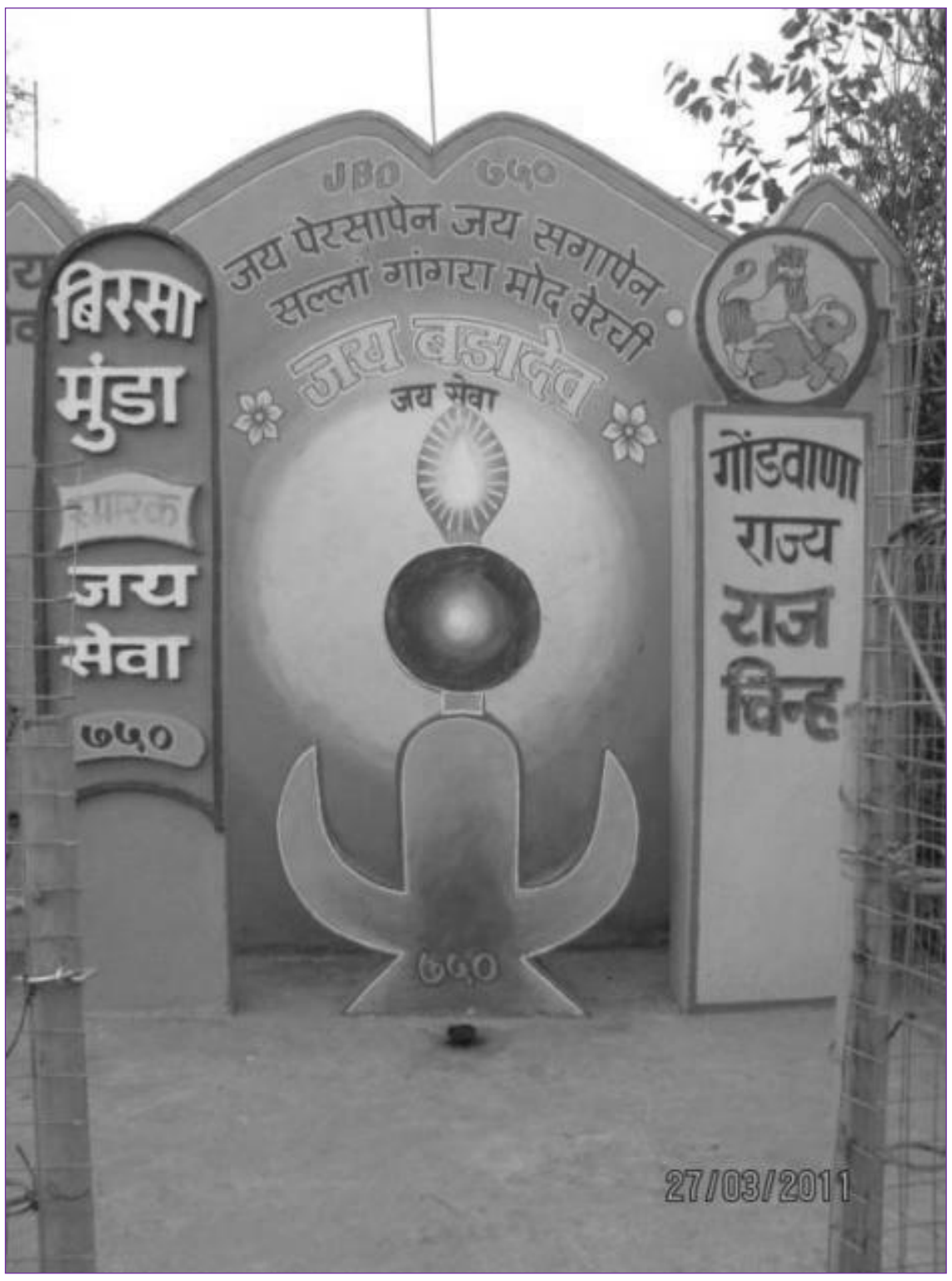

Figure 1: The religious symbol of the Gonds.

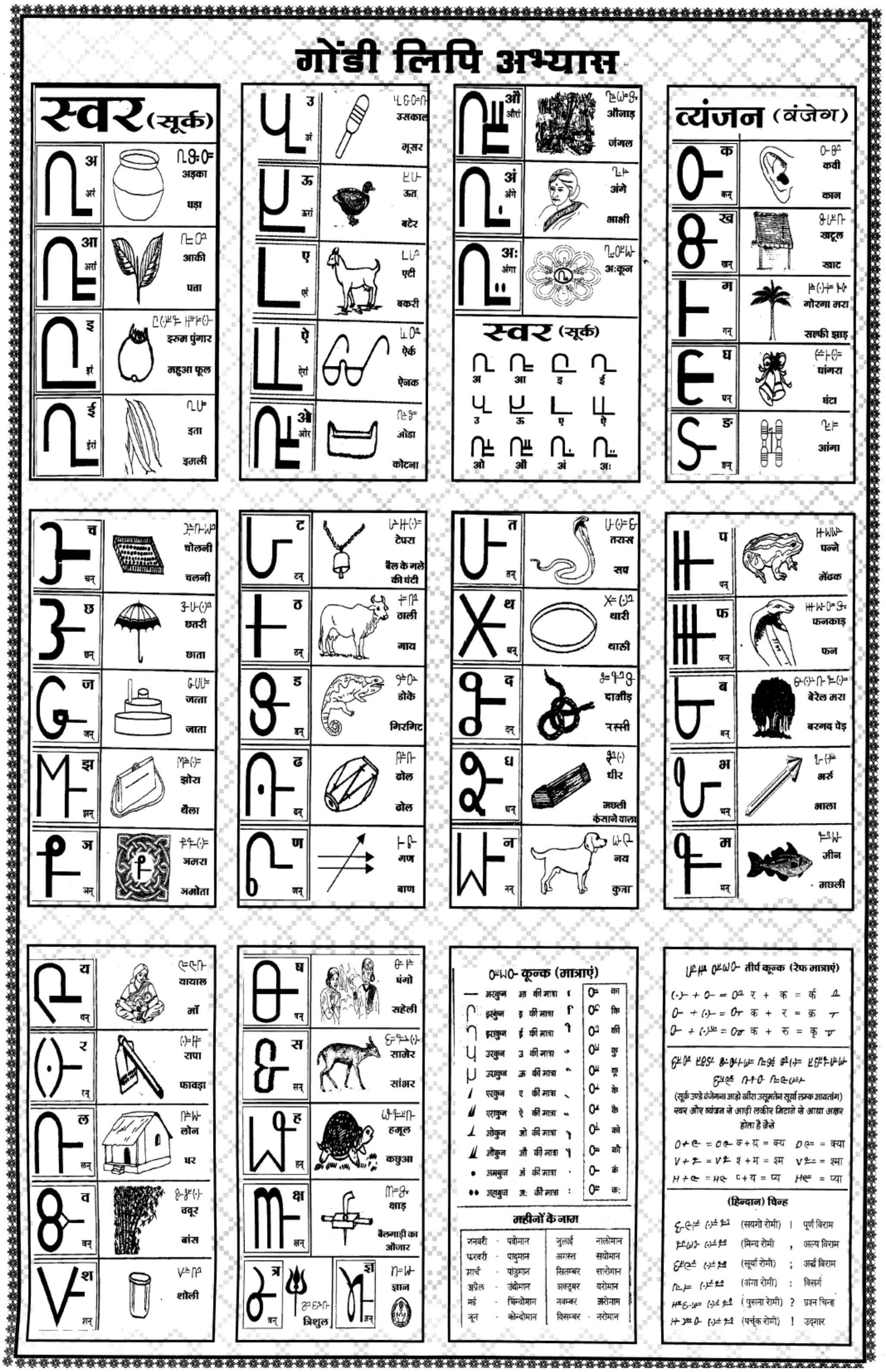

Figure 2: A description of the Gondi script.

Table 2: Days of the week in Gondi and in other languages.

| Day of the week | Name in English | Name in Hindi | Name in Telugu | Name in Gondi |
|---|---|---|---|---|
| 1 | Sunday | Ravi vaar | Aadhi Vaaramu | Purva net |
| 2 | Monday | Som vaar | Soma Vaaramu | Nalla net |
| 3 | Tuesday | Mangal vaar | Mangala Vaaramu | Surka net |
| 4 | Wednesday | Budh vaar | Budha Vaaramu | Surva net |
| 5 | Thursday | Guru vaar | Guru Vaaramu | Mudha net |
| 6 | Friday | Shukra vaar | Sukra Vaaramu | Nilu net |
| 7 | Saturday | Shani vaar | Seni Vaaramu | Aaru net |

Stylistically, the script differs significantly from other Indian scripts including Indus, Devnagari and the Dravidian group of languages, although it includes signs for consonants (such as a deep N)[3] which are no longer used in Hindi but are common in Marathi. We obtained three different calendars from the three different regions of Andhra Pradesh, Maharashtra and Madhya Pradesh. The calendar from Andhra Pradesh was wholly in Telugu while the one from Maharashtra was in Gondi and Marthi, and the one from Madhya Pradesh was in Hindi and Gondi. The numerals used in these calendars are given in Table 4. In the listing of months and days, the calendars of Madhya Pradesh and Maharashtra agree in detail (except for some obvious printing errors), but they differ significantly in the signs for the numbers 5 and 6. The Gonds have separate names for the numbers 1 to 10; after that they use the system of tens first followed by the numerals (i.e. 10 and 3 for 13, not 3 and 10 which is used in Hindi, for example).

## 2.6 The Myths of the Gonds

Deogaonkar (2007: 123-130) has briefly discussed the myths and folk literature of the Gonds, while Mehta (1984: 167-306) has discussed their myths and subtle regional differences in detail. Interestingly, all the recorded myths are related to terrestrial aspects, and stories of Great Floods and the virgin birth of the goddess are very common. Mehta (1984:181) considers the Gond hero *Lingo* to be the equivalent of Moses of the Jews who, with the mercy of the *Bada Deo*, his wife and *Gangudevi* the Great Goddess, freed them from the curse of captivity and led them to freedom. According to Mehta (1984: 37), the *Bada Deo* (also called *Pen*) is synonymous with *Mahadeo* and *Shiva*. Mehta (1984: 38) also refers to the *Bada Deo's* wife as *Parvati*, but this association is not obvious. The image of the *Bada Deo* differs from the conventional image of *Shiva* in many significant ways. For one, he is a creator who, after having initially banished the Gonds for bad behaviour turned around to assist them to the extent of taking on rivalry with *Indra* to create the Gonds (Mehta, 1984: 180). The *Bada Deo* also assists *Lingo* in a variety of ways.

It is interesting that in their analyses of Gond myths and beliefs neither Deogaonkar (2007) nor Mehta (1984) makes any reference to astronomical or cosmogonical ideas. The closest they come are in their discussions of the Great Floods, or the inability of the "… Sun, Moon and Stars to assist Lingo in locating the banished Gonds." (Mehta, 1984: 184). They take the terrestrial world to have been in existence forever, their land being the land of seven mountains and twelve hills (Mehta, 1984: 178). They also suggest that the Earth is held on the head of *Patar Shek* (Mehta, 1984: 187). The Gond calendar from Andhra Pradesh (see Note 2) states that according to the Gonds

> The gift of nature, which gives astronomical, magnetic and gravitational pull makes the Earth move from right to left, that is, in an anticlockwise direction. (our English translation).

Table 3: Months of the year in English and in Gondi.

| Day of the week | Name in English | Name in Gondi |
|---|---|---|
| 1 | January | Pado man |
| 2 | February | Padu man |
| 3 | March | Pandu man |
| 4 | April | Undo man |
| 5 | May | Chindo man |
| 6 | June | Kondo man |
| 7 | July | Naalo man |
| 8 | August | Sayo man |
| 9 | September | Saro man |
| 10 | October | Yero man |
| 11 | November | Aro man |
| 12 | December | Naro man |

Beyond this, there are no records of Gond astronomical ideas.

However, since they held sway over large tracts of land and administered them, they must have had calendrical and other time-keeping systems. Such systems are most often rooted in astronomy, and hence observational astronomy must have been an important aspect of the science of the Gond people. Since they were never integrated into the dominant cultural and population groups of India until recently, their knowledge presumably contains the seeds of an independently-developed perspective of the Universe. In order to understand this, we studied the astronomical knowledge of the Gond people. In this study we focussed solely on understanding their astronomical traditions and ideas. As we have observed in our study, the limited description of Gond mythology is only a partial truth, and the skies form an integral part of their life—as would be expected.

Table 4: Numerals in the Gond script, and in Maharashtra and Madhya Pradesh.

| Number | Name | Style in Maharashtra | Style in Madhya Pradesh |
|---|---|---|---|
| 1 | Undi | 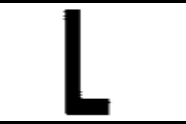 | 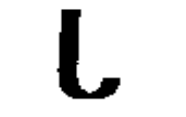 |
| 2 | Rand | 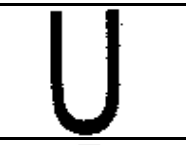 | 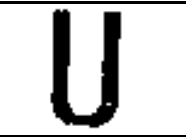 |
| 3 | Munda | 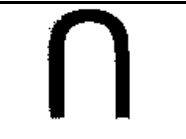 | 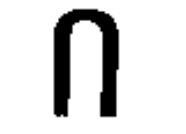 |
| 4 | Nalung | 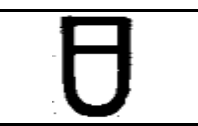 | 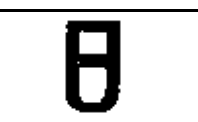 |
| 5 | Sayyung | 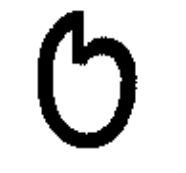 | 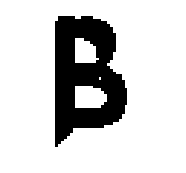 |
| 6 | Sarung | 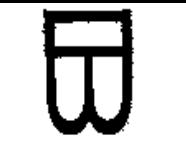 | 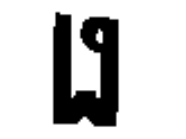 |
| 7 | Yerung | 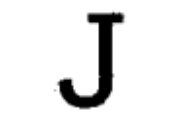 | 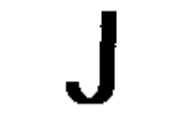 |
| 8 | Arung | 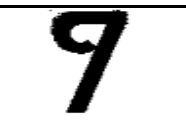 | 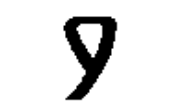 |
| 9 | Narung | 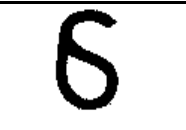 | 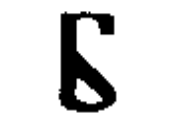 |
| 10 | Pad | 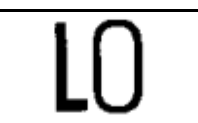 | 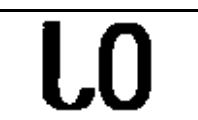 |

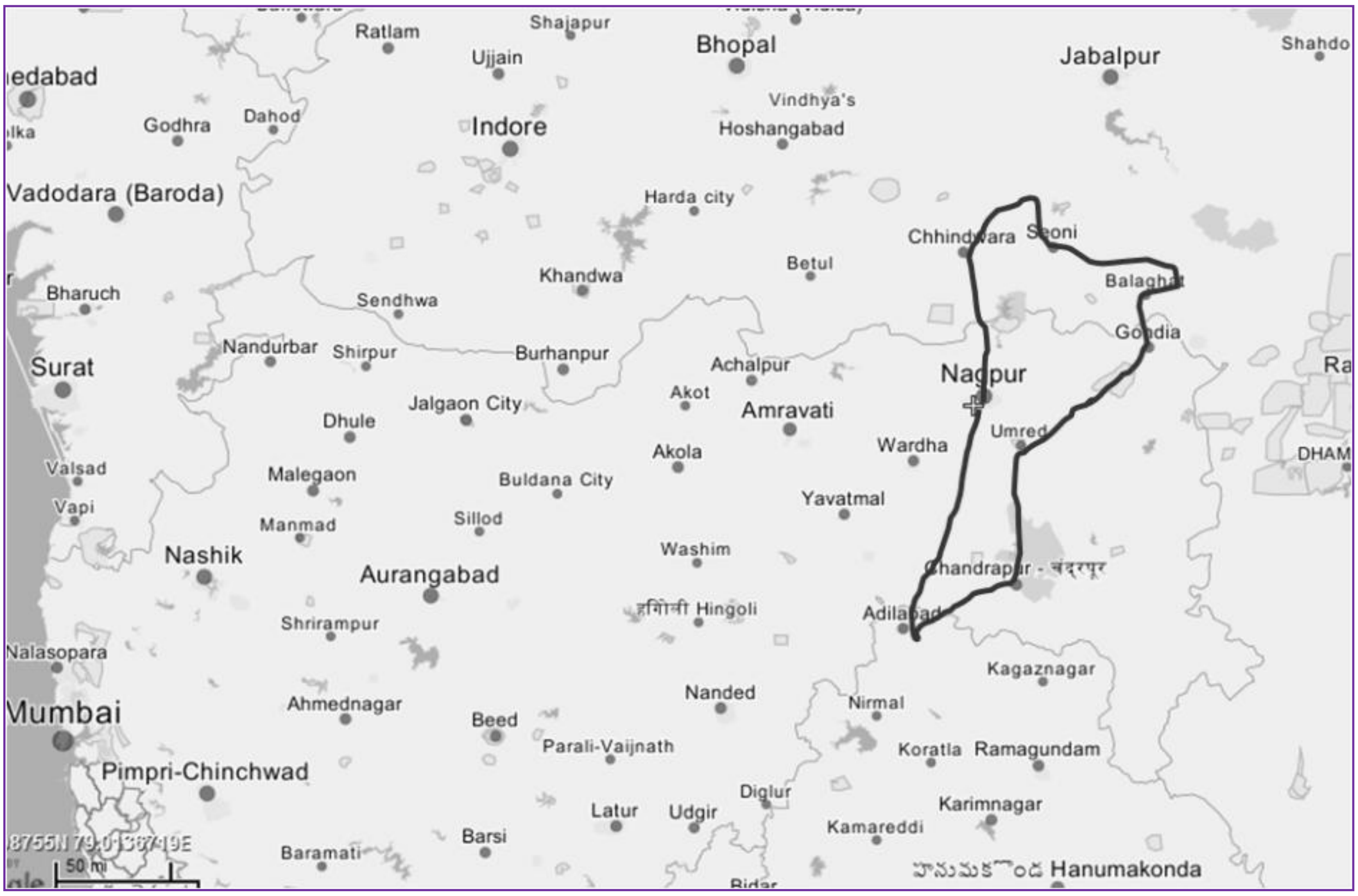

Figure 3: Map of our travels and villages visited by us.

## 3 THE PRESENT STUDY

From 25 to 31 March 2011 we visited 15 Gond villages spread over an area of 2000 $km^2$ around the Nagpur region in the states of Maharashtra, Andhra Pradesh and Madhya Pradesh, covering six of the seven regions mentioned in Section 2.2. In Figure 3 we show the area surveyed and the path that we followed. All 15 villages lay within the latitude range 19.5° and 21.8° North.

Details of the villages visited, persons contacted and the astronomical knowledge they supplied are given below. In order to ensure that all of us were talking about the same region of the sky, we carried a laptop and a LCD projector and whenever it was necessary we projected an image of the sky on the walls so that constellation identifications could be confirmed. While this approach was successful on most occasions, there were villages where the audience could not fully identify with the projected sky. The 15 villages visited by us are now discussed individually.

### 1: Karambi

State: Maharashtra

Location: About 22 km from the Shankarpur village of Chimur Tehsil in Chandrapur district, 100 km east of Nagpur, near Nagbhid.

Date of visit: 25 January 2011

Person contacted: Kirtivat Tivalsingh Atram.

Astronomical knowledge: They know the Belt of Orion as *Tipan* and the Pole Star as *Lagni Sukum* or ‘the bright one’. *Sukum* means ‘star’ and *Lagni* means ‘the one that shines’. The Sun and the Moon are called *Lingo* and *Jango* respectively.

### 2: Nimni

State: Maharashtra

Location: Post Jamkola, Taluka Zari, District Yeotmal. Via the Pandharkawada-Ghonsa-Wani road, 25 km from Pandharkawada. The village has 60 to 65 Gond houses. The other houses belong to Kolam tribesmen.

Date of visit: 25 March 2011

Persons contacted and their ages: Mahadeo Anandrao Kudmethe (56) and Sanjay Masram (28).

Astronomical knowledge: A star is called *Sukum*. *Saptarshi* (*Katul* and *Kalher*), *Samdur* (sea) a group of four stars in the shape of a quadrilateral (probably Auriga), comes overhead at 4 am, rains arrive and farming begins. They know the Belt of Orion as *Tipan*. 4 am, *pahili chandani*, heralds the beginning of the working day. Evening is known as *dohan chan-* *dani* and is the time to milk the cows. The Sun is *Lingo* and the Moon is *Jango*. Comets are called *Jhadani* (in Marathi), which means a broom. Comets are the weapon of the Great God *Bhimal Pen* called *Bhimal-Saat*, and he uses them to cleanse the sins of the world so they are thought of as good omens. The prevailing burial practice is in the North-South direction, but a special undercut is made near the head so that top soil is not dug up near it and the head is slipped inside. Our informants know that the length of the day changes during the year. They also know the four cardinal directions as *Silalin* (East), *Farayin* (West), *Kalvada* (North) and *Talvada* (South). They are familiar with the southward and northward movement of the Sun in the course of the year and its relation to the seasons.

### 3: Matharjun

State: Maharashtra

Location: Post Matharjun, Taluka Zari Jamni, District Yeotmal. On the Pandharkawda-Shibla-Matharjun road, and the village is 28 km from Pandharkawda. There are about 170 Gond houses.

Date of visit: 25 March 2011

Persons contacted: Shyamrao Aatram (65), Gopalrao Maraskolhe (65), Punaji Madavi (70), Deorao Dongaru Madavi (70) and Karu Keshav Madavi (80).

Astronomical knowledge: *Shukra* (a canonical star that appears at sunset in the east) is called *Jevan-sukum*. They know *Tipan* (the Belt of Orion), and next to it are *Medi* (a star pattern where there is a bright star in the centre and other stars in a circle appearing like the set up for crush grains etc. using a bullock, locally known as *Khala*; but the constellation association is unclear) and *Tiva* (meaning a stool on which a farmer stands to thresh the grains by dropping them on the ground in the wind), which is identified with a constellation pattern just south of Sirius. They have rudimentary knowledge of how to predict the seasons by the presence of different star formations. They know a constellation called *Katul* (meaning a cot) as part of the constellation that we recognise as *Saptarshi* (or the Big Dipper). The first four stars that make the cup of *Saptarshi* are *Katul*, which is imagined as a cot with four legs made of the following precious metals: gold, silver, inferior silver and copper. The last three stars of *Saptarshi* are called *Kalher* (meaning thieves) and represent three thieves who want to steal the cot when the old lady falls asleep. Hence the old lady never sleeps (possibly indicating that the constellation never sets—as was the case in this region of India until 1000 BC, when *Saptarshi* was partly circumpolar). They know another constellation that they called *Samdur* (a quadrilateral, probably Auriga) which indicates the arrival of the monsoon season. A constellation they call *Kotela* (meaning the tool shaped like a cricket bat that is used to beat the grain out of the husk), most probably is the Pleiades (but it could be Taurus). They also identify *Koropadera* (a tool used to make buttermilk) but its modern equivalent was not identified by us. The constellation of *Koropadera* is a good omen. They also know *Michu* (the Scorpion) which is the same as the modern constellation of Scorpius. They know the names of the four cardinal directions: East is *Shilain*, West is *Farain*, North is *Kalwada* and South is *Talwada*. While visiting here we spotted a rainbow and they called it *Kamarpatta*. The Milky Way is *Sagur*, or a road. They were aware of Gondi numbers from 1 to 7 (8 onwards is Marathi). They buried the dead in a North-South direction, with the head to the north. A comet is called *Bhimalsaat*, and a shooting star is *Sukum Pelkta* (star excreta).

### 4: Kesalapur

State: Andhra Pradesh

Location: Post Kesalapur, Mandal Inderveli, Taluka Utanoor, District Adilabad. Via Adilabad, Gudi Hatnoor and Mutnoor. About 35 km from Adilabad.

Date of visit: 26 March 2011

Persons contacted: Urvetta Chinnu (100), Mesram Laxman (45), Mesram Venkatrao (45, the *Patel* of the village) and Todsam Ghanshyam (60).

Astronomical knowledge: This village had the most detailed memory of astronomy. They gave us the names of the 12 months of the year and the *adhika maas* (intercalary month). They know *Saptarshi* as *Katul* (old lady's cot) and *Kalir* (thieves), and the Milky Way as the 'Path of the Animals' (*Dhor Sari Marg*). The location of *Saptarshi* at sunset is used for calendrical purposes. They know that there are three star combinations for three seasons. Season 1 has the constellation of *Murda* in which they identify not only the body, but also the complete funeral procession of stars in this order: *Duf* (the drum bearer), *Shika* (the procession leader), *Murda* (the dead body) and a group of 4 *Ladavya* (weeping women). *Murda* rises around 11 pm early in this season. This star combination is to the north (east?) of Scorpius (whose bite produces the *Murda*) and extends across the sky. They know a constellation *Purad* or *Hola*, a bird and its two eggs, probably stars below Sirius. They know *Pahat sukum* (meaning 'morning star'), which they identified as Alpha Aquila when they saw the projected star chart. Season 2 has the months of *Budbhavai*, *Akhadi* and *Divali*. During this time you see *Samdur* (Sagittarius?), *Tipan* (Orion) and *Topali* (Lepus). Season 3 is from *Kartik* to *Maho* and the prevailing constellations are *Medi* (rising between 9 and 10 pm), the central star of *Khala* (a grain-crushing device), *Tiva* (a stool) and *Kotela* (a bat). In June at sunrise they can see *Tipan* (the Belt of Orion), *Topli* (Gamma and Beta Orionis), *Samudar* (Alpha, Beta, Gamma and Kappa Cassiopeia), *Medi* (Taurus) and *Tiva* (Canis Major)—confirmed by our projection of the night sky. They know that the monsoon arrives when *Tipan* appears at sunset. They know the Orion sequence with *Kotela* (a bat for thrashing grain) identified with modern-day *Lepus,* and can identify a basket (*tokali*). They know of comets as *Kayshar* (a broom—the Pleiades) and shooting stars as star excreta (*Sukir Pelkta*). A month is called *Vata*, and extends from New Moon to New Moon. The northern direction is considered inauspicious.

### 5: Kharmat

State: Maharashtra

Location: Post Kharmat, taluka Pombhurana, District Chandrapur, via Rajura, Kothari, Gondpimpari, Kharmat, 57 km from Rajura (GPS location 19° 48′ 6.5″ N and 79° 39′ 1.3″E).

Date of visit: 26 March 2011

Persons contacted: Sitaram Kisan Madavi (75), Sambhashiv Shivarama Madavi (62) and eleven other members of the Gond tribe.

Astronomical knowledge: They know *Katul*, *Jevan Sukum* (the first star seen in the east at night, indicating dinner time). Similarly there is *Pahat Sukum* (the last star to be seen in the west at sunrise), *Naagarda* (Orion), *Sagur* (the Milky Way) seen at midnight, *Irukmara* (a star seen at 3 am) which indicates the season for collecting *Mahua* (*Madhuca longifolia)*. They know a constellation *Kutpari* or *Mogari* which they identify with the modern Pleiades, and refer to a comet as *Kayshar* (a broom) and a shooting star as *Chandani Pelkta*. They know the 12 months of a year; the intercalary month after completing three years; and can count to 7 in Gondi and after that in Marathi. They call the Sun *Purbaal* and the Moon *Nalend*, and East is *Sukra*l. Their burials are in a North-South direction. They still remember a major flood that occurred about 3,500 years ago. At the end of the flood crows brought soil to create the Earth, and hence the crow is worshipped. They 'predict' the arrival of the monsoon by seeing clouds in the west. New Year *Punal (Nava) Saal* starts at *Gudi Padwa.*

### 6: Wamanpalli

State: Maharashtra

Location: Post Lathi, Taluqa Gondpimpari, District Chandrapur. Via Gondpimpri, Dhaba, Sonapur. 35 km from Gondpimpri on the banks of the Vardha (across the river from Andhra).

Date of visit: 27 March 2011

Persons contacted: Sainath Kodape (28, the Village *Sarpanch*), Kawdu Raju Gadaam (70), Urkudabai Sukru Veladi (65), Bhiva Kondu Talandi (66), Ganpat Dharma Sedmake (78), Urkuda Paika Sedmake (60) and Gopala Tanu Madavi.

Astronomical knowledge: They know that stars are called *Sukum*. They know the constellation termed *Tipan* by others by a different name, and call it *Naagarada* (which means 'plough') and identify it with the modern-day Belt of Orion. They know *Saptarshi* with an imagery and mythology that is similar to other regions namely, the cot is called *Sedona Katul*, and the three thieves are called *Muvir Kaler*. They correctly identify the legs of the cot. In the sky they can also see *Irukna Mara* (the tree of *Mahua*, *Madhuca longifolia*), *Pahat Sukir* (the Morning Star), *Jevan Sukir* (the Evening Star), *Dhruva* (clearly an after-thought and addition since the Gonds have no word or reference to it, and many other villages denied its existence), *Kutpari* (the Pleiades), *Kayshar* (the broom = comet) and the Milky Way as *Pandhan* or *Sagur* or *Murana Sagur* (the path of animals). Their list of months is the same as the generic list. The Moon is called *Nalen*. From New Moon to Full Moon is called *Avas*, and from Full Moon to New Moon is *Punvi*. The lunar calendar is followed, and they recognise the intercalary month. A shooting star is *Sukir Pelkta* (star excreta). Human burial is North-South. The burial itself should be far from home, but a memorial stone can be set close to home, with a terracota or wooden horse that is worshipped for generations when convenient. None of these memorial stones we saw were more than a hundred years old, indicating that ancestor worship is forgotten after a generation or so.

### 7: Khadaki

State: Maharashtra

Location: Post Mendha, Taluka Nagbhid, District Chandrapur. Via Nagbhid, on the Nagbhid-Mendha-Khadaki path. On the Nagpur-Brahmapuri Road about 12 km from Brahmapuri.

Date of visit: 27 March 2011

Persons contacted: Sheshrao Mansaran Naitam (38, *Up-sarpanch*), Narayan Bisan Madavi (70) and Barikrao Sitaram Naitam (65).

Astronomical knowledge: They know *Shukra* as a generic evening star, *Tipan* (Orion), *Mongari* (the bat = Pleiades), *Jevan Chandani* (the first star of evening), *Saptarshi*, Scorpius (*Vinchu*), a comet as *Kayshar* (which is vaguely regarded as a portender of bad luck), the Milky Way (*aakash ganga*) and a meteor shower (*ulka*). They believe that the world moves counterclockwise as do the planets, whirlwinds and whirlpools, and the oil-extracting bull-run grinding device common in India. Their burials are in the North-South direction to the East of the village (but this latter choice seems to have been made more out of local geographical necessity rather than some custom). Burials include personal utensils and other belongings. Now-a-days they include dolls made from edible flour. The grave of an old man is marked by a vertical stone, while other graves are left unmarked. They worship their ancestors in the form of horses. They recognise the intercalary month.

### 8: Yelodi

State: Maharashtra

Location: Post Dhabe Pawani, Taluka Arjuni Morgaon District Gondia. Via Brahmapuri-Wadse, Arjuni Morgaon-Navegaon Bhandh and Dhabepawani, on the Dabepawani-Chikalgad Road, 5 km from Dhabe Pawani and 28 km from Arjuni.

Date of visit: 28 March 2011

Persons contacted: Jairam Manku Salame (75), Kaaru Devsu Duge (65), Charandas Nagaru Kumare (60), Pandhani Istari Walke (70), Jagan Mansaram Walke (70), Baliram Dhondu Ghumake (70), Sadashiv Laxman Kokote (65), Govinda Bakshi Uike (80), Tukaram Madku Karpate (67) and Goma Ghegu Alone (60).

Astronomical knowledge: They have a vague idea of *Jevan Chadani* (the Evening Star), *Pahat Tara* (the Morning Star), *Saptarshi* (*Katul* and *Kalher*), Orion (*Nangal*, visible in the east every day), *Thengari* (the bat = the Pleiades) and *Topli* (Lepus). They know *Sagar* (the Milky Way), and the Moon as *Nanleg* and the Sun as *Bera*. Their burials are oriented East-West, with the head to the East.

### 9: Zashinagar

State: Maharashtra

Location: Post Palasgaon chutia, Taluka Arjuni Morgaon, District Gondia via Navegaon bandha and the Dhabepaulani-Chichgad Road, 16 km from Navegaon.

Date of visit: 28 March 2011

Persons contacted: Antaram Modu Bhogare (78) and Sitaram Chamru Hodi.

Astronomical knowledge: They know the Moon as *Nalen* and the Sun and *Vera*. They know *Saptashi* as *Sedona Katul* and *Kaler*. They know that the first leg of *Katul* is made of gold. They know the early morning star as *Viya Huko* (‘Huko’ means ‘star’). They know the first star of the night as *Jevan Sakun*. They know *Nangal*, and they refer to the Milky Way as *Hari*, or the ‘road’. They know about *Bohari* (*Kayasur*) or *Jhadani* (but they could not point one out). A shooting star is called *Huko Pelkta*. They know the names of each month. They have heard about the Gondi *lipi* (script) but have no idea what it is like. They know of the equinox, and they bury the dead East-West with head to the East.

### 10: Mohagaon

State: Maharashtra

Location: Post Supalipa, Taluka Aamgaon, District Gondia via Gondia, Dohegaon, Adasi, Gudma, Sitepar and Mohogaon.

Date of visit: 28 March 2011

Persons contacted: Ramlalji Uikey (69), Beniram Yadu Uikey (55) and Nimalabai Uiley (50).

Astronomical knowledge: They know *Nangar* (Orion’s Belt) and can point it out. They know *Pahat Sukir* (a Morning Star that rises every morning at 4 am). They know *Saptarshi*. They do not know the Pole Star, Polaris. They know *Kotela* (*Taurus* or the Pleaides) and *Topli* (Lepus). They call the Milky Way *Sagarpath*. The Sun is *Din* and the Moon is *Chandal*. They know names of the months. A comet is called *Kaysaar*. They know that a glow called *Kondor* appears around the Moon, and if it is close to the Moon the rain is far away but if it is far from the Moon then the rain is nearby. They bury the dead in a North-South direction, and they sometimes include burial goods such as clay pots for use in the afterlife.

### 11: Kaweli

State: Madhya Pradesh

Location: Post Chalisbodi, Block Parswada, Tehsil Baihar, District Balaghat. Via Balaghat-Banjari-Kanatola-Kaweli. Banjari is 21 km from Balaghat on the Baihar Road. From Banjeri to Kaweri is 9 km.

Date of visit: 29 March 2011

Persons contacted: Sohansingh Bilaising Uiykey (41), Munnalal Zarusing Bhalavi (70) and Himmatsing Mohan Uikey (40).

Astronomical knowledge: They know the Morning Star rising at 4 am and the Evening Star. They say that *Nangar* (the plough) rises every evening. They know *Katul* and *Kalhad* (*Saptarshi*). They know the Pleaides as *Kayshar* (*Bahari*), the broom. They know *Jewan Tara* (a late evening star). They think the Pole Star rises at 4 am. They know *Nangar* but think it rises every evening or morning. They know *Saptarshi*. They refer to a shooting star (meteor) as star excreta, and have heard of comets but do not know much about them. They can identify the months of the year. They know that one month runs from New Moon to New Moon and every third year is a ‘*Dhonda*’ year when an intercalary month is added; no marriage can occur during this month. They know of the glow around the Moon and can identify it. The Sun is *Din* and the Moon is *Chandal*. They have not heard of eclipses. They count from one to seven in Gondi, and they bury their dead with the body aligned North-South.

### 12: Chalisbodi

State: Madhya Pradesh

Location: Post Chalisbodi, Block Parswada,

Thesil Baihar, District Balaghat. Via Balaghat-Banjari-Kanatola-Kaweli. Banjari is 25 km from Balaghat on the Baihar Road, and the distance to Kaweli is 4 km. It is 35 km from Balaghat and 13 km from Banjari.

Date of visit: 29 March 2011

Persons contacted: Gorelal Madavi (60), Mohanlal Tekam (55), Radheshyam Warkale (52), Mohparsingh Markam (35) and Ramsing Tekam (34).

Astronomical knowledge: They know of the Pole Star that is seen every day. They know *Sedona* (old woman's) *katul* (cot) and *Kalhad* (*Mund kalhed*, i.e. thieves). They know *Nangar*, that is like a plough, and *Kotela*. In addition, they know of *Purad* and *Mes* (a bird and its egg) as stars east of Sirius. The story goes that the man in Orion throws stones in the form of the Pleiades so that they will fall on the bird and kill it. What the story does not record is if he was successful. They know the Evening and Morning Stars. They can count to 7 in Gondi, and they claim that *Aimdi* is 10 and *Padi* or *Padivakati* is 100. They count 12 months of a year, and the leap month. They bury their dead facing North-South. They know of comets as *Jhadu* and shooting stars as stellar excreta. They refer to a rainbow as *Gulel*, the bow of a bow and arrow. They know of the glow around the Moon. They know *Pada din* (increasing day), and *Chirdur din* (decreasing day). To them the Milky Way is *Sadak*, and they have heard of the Gondi script.

### 13: Kopariya

State: Madhya Pradesh

Location: Post Ramnagar, Block Mohagaon, Thesil Mandala, Julla Mandala.

Date of visit: 30 March 2011

Person contacted: Shivsingh Charusing Parateti (70).

Astronomical knowledge: They know *Nangir* (Orion), can point it out and know that it rises around 8 pm in April and brings rain. They know *Dhruva Tara*. They know *Mangal Tara*, which is the morning star. They know *Poyi* (a noble man), his wife (*poyatar*) and his *kotwal* as the three stars that form the tail of *Saptarshi*. The *Kutil* is the path of *salveshan* and the three approach it for their personal salvation after doing good deeds on Earth. They know the Morning and Evening Star. They have heard of Scorpius, and know of the Pleiades as *Kotela*. They refer to the Sun and the Moon as *Dinad* and *Chandal* respectively. They know comets as *Baahari* (the broom) and shooting stars as *Tara Uruganta*. They know of the glow around the Moon and its interpretation. A month goes from New Moon to New Moon, but in contrast they claim that each month has exactly 30 days; they do not know that an intercalary month is added after three years. They bury their dead oriented North-South.

### 14: Sailakota

State: Madhya Pradesh

Location: Post Kanhiwada, Block Seoni and district Seoni. Salaikota is 26 km from Sivani.

Date of visit: 31 March 2011

Person contacted: Sabalsingh Kaureti (72).

Astronomical knowledge: They know about *Saptarshi* but are confused about the story. They know *Bahri* and pointed it out in the sky as the Pleiades. They know of the Sun as *Din* and the Moon as *Chandal*. They know the Milky Way, and believe that shooting stars occur when souls fall back to Earth. They refer to a comet as a broom (*bahari*). They know of the glow around the Moon and can interpret it correctly in terms of its relation to rain. They do not know about eclipses. They can count a little in Gondi and can recite the months. They know that the Gondi script probably exists. They bury their dead in a North-South direction.

### 15: Lodha

State: Maharashtra

Location: Post Karwahi, Tehsil Ramtek, District Nagpur. Via Manegaon tek, Karwahi, Lodha, Pindkepar. 11 km from Manegaontek.

Date of visit: 31 March

Persons contacted: Munsi Saddi Bhalavi (75) and Parasram Munsi Bhalavi (45).

Astronomical knowledge: They know the Morning Star and the Evening Star, *Saptarshi*, *Kaysar* (the Pleiades), Scorpius (?), *Purad* and the glow around the Moon. They refer to shooting stars as stellar excreta, the Moon as *Chandal*, the Sun as *Suryal* or *Din* and the Milky Way as *Sari*. They know the months of the year, about leap years and about the numbering system. They bury their dead North-South (where South is termed *Rakshas Disha*).

## 4 ANALYSIS OF THE OBSERVATIONS

In Table 5 we list major aspects of astronomy known to the Gonds. In most cases, the information was corroborated from more than one village, although in some cases the precise detail of the name varied as a result of local linguistic differences. In April 2012 a select group of 23 villagers from the three districts of Adilabad, Yeotmal and Chandrapur were invited to the Raman Science Centre in Nagpur, and asked to explain the night sky in the planetarium.[4] The associations discussed below therefore represent accurate identifications.

In Table 6 we list the villages in which we were told the same or largely similar stories or identification names of various objects.

Table 5: A list of the major astronomical ideas of the Gonds.

| | Standard Terms | Local Names | Description |
|---|---|---|---|
| | | Sun, Moon, etc. | |
| 1 | Sun | *Lingo*, *Purbaal*, *Bera*, *Vera*, *Din*, *Dinad*, *Suryal* | There is no knowledge of solar eclipses. |
| 2 | Moon | *Jango*, *Chandal, Nalend* or *Nalen*. New Moon to Full Moon is *Avas*, and Full Moon to New Moon is *Punvi*. | There is no knowledge of lunar eclipses. |
| 3 | Glow around the Moon | *Kondor* | Often a glow is seen around the Moon. If the glow is close to the Moon, the rain is far away while if it is far from the Moon the rain is expected. |
| 4 | Duration of the month | The month runs from New Moon to New Moon. | There is no long-term calendar. Every 3rd year has a 13th month for solar-lunar synchronisation. New Year begins at *Gudi Padwa* though in earlier times the dates were probably different. |
| 5 | Months of the year, and the leap month | *Vata* (month), *Punal (Nava) Saal* (New Year) is on *Gudi Padwa* though older practice was different. A leap month is called *Dhonda*. Increasing length of the day is called *Pada din* and decreasing length of the day is called *Chirdur din*. | January = *Pus*, February = *Maho*, March = *Ghuradi* (*Umadi Amavasya* marks the New Year), April = *Chaita*, May = *Bhaavai*, June = *Bud Bhaaavi*, July = *Aakhadi*, August = *Pora*, September = *Akarpur*, October = *Divali*, November = *Kaartika*, December = *Sati*. At the end of every three years the New Year is delayed by 1 month by adding a *Ghoda*. A month runs from New Moon to New Moon and no calculations are done. Sometimes a *tithi*, (Lunar Mansion) particularly *Amavasya*, can extend to 2 days. Long-term memory does not go beyond 3 years. |
| 6 | Directions | *Silalin* (East), *Farayin* (West), *Kalvada* (North) and *Talvada* (South).. | Directions are important to the Gonds largely for burial rituals. We did not come across any evidence where they use the stars for navigation. In one village they knew of the northward and southward movement of Sun and its relation to the seasons. |
| 7 | Burial practices | | Burials are most often aligned North-South, with the head to the South, since "… bad people live in and come from the North." Some people practised equinoctial East-West burial, with the head to the East. In the village of Nimani we were told that while the body is laid straight, the head is put under the solid earth by scooping out more earth on that side. There were few reports of burial of goods with the body. A grave was typically about 3 feet deep, i.e waist deep. |
| 8 | A rainbow | *Kamarpatta*, *Gulel* | It is also called the bow of a bow and arrow. |
| | | Stars | |
| 9 | A star | *Sukum*, *Sukir*, *Huko*, *Tara*, the adjective *Lagni* ('bright') is also used. *Sukra* is also used. | These are generic names for all stars. These terms are also used to describe diffuse moonlight or starlight. *Chandani* is also a generic name for starlight or moonlight |
| 10 | The Morning and Evening Stars (The planet Venus). | *Jevan Tara* ('dinner star', is also called *Shukra Tara* or *Pahili* [first] *chandani*), and *pahat sukom* (*sukur*), (star of early morning), *Shukra*, *Mangal Tara*. The Evening Star is also called *Dohan Chandani* and indicates the time to milk the cow.<br>One group (from the Matar-jun region) was categorical that this star rose with the Sun either in the morning or the evening. | *Jevan* (meal) is a generic star that rises every evening in the east, indicating dinner time. *Javen*/*Pahili* etc. *tara* is a generic early morning star that is overhead at 4 am(!) indicating the time to start working.<br>The recognition that *Jevan Sukum*, the Evening Star, and *Pahat Sukum*, the Morning Star, is the same *Sukum* indicates a knowledge of planets, or transient stars. However, with the exception of Venus, the Gonds are not aware of any other transient stars. |
| 11 | Comets | *Jhadani*, *Bhimal Saat*, *Kay-shar*, *Jhadu*, *Bahari* | A comet is believed to be the sword-like weapon of the gods, and is considered a good omen in that the gods are protecting humans by cleaning up the mess that was created by bad events, either by killing evil (using the sword) or sweeping away the evil (with a broom). |
| 12 | Shooting stars | *Ulka*, *Sukum Pelkta*, *Sukir Pelkta*, *Huko Pelkat*, *Tara Urungta*. | In general shooting stars (meteors) are called excreta of stars, or are thought of as souls that are falling from their holy places in the sky. |
| 13 | Milky Way | *Dhor Sari*, *Rasta*, *Sagur*, *Murana Sagur*, *Marg*, *Pandhan*, *Hari* ('the road'), *Sadak* | The Milky Way is known as the great path of animal migration. |
| 14 | The Pole Star | *Dhruva Tara*, *Mout Tara* | Polaris was reported in three villages, using a Sanskrit name, which suggests that it is a later addition. *Mout Tara* is the umbilical star. |
| | | Constellations | |
| 15 | *Saptarshi* | *Sedona* (old lady's) *Katul* (cot) and *Kalher* or *Kalhed* (thieves), [*muvir* = three] *Kal-hed*, *Kalir*. *Buddhi chi khat* and *chor*. In another story the | It is believed that the first four stars of the *Saptarhi* form the bed of an old lady and that the legs of this bed consist of gold, silver, inferior silver and copper, in an anti-clockwise direction from the star of contact to the trailing three that form the three thieves who are trying to steal the bed. In turn they keep the old lady from falling |
| | | three thieves are replaced by *Poyi* (a noble man), *poy-atar* (his wife) and *Kotwal* (his assistant) going towards their salvation. | asleep. It is believed that if the old lady sleeps, i.e. if *Saptarshi* sets, the Earth will come to an end. This refers to the circumpolar nature of *Saptarshi*.<br>*Saptarshi* is the primary reference point from which all constellations are located. |
| 16 | Auriga | *Samdur* | This is visible in the last week of May at 4 am indicates the arrival of the monsoon season. The constellation is overhead in early July at 4 am in the morning. If Auriga is bright at that time, it is assumed that the monsoon will be good and the Gonds sow water-demanding crops like cotton, but if Auriga is dull they assume that the monsoon will be weak and as a result they sow crops that will need less water. |
| 17 | Orion and its Belt | *Tipan*, *Naagarda*, *Nangar*, *Nangir*, *Nangal*. | The Belt of Orion is called *Tipan* (3 stars) while along with the sword of Orion, it is called *Naagarda*, which is like a plough. With Taurus, the eastern shoulder of Orion, Lepus and Sirius, it refers to farming activities. The arrival of *Tipan* in the early night sky therefore is an indication of the arrival of the farming season. |
| 18 | Sirius region | *Topli* | The basket is indicative of the basket of seeds which is used for sowing in the fields ploughed by *Tipan*. |
| 19 | Taurus | *Medi, Kotela* | |
| 20 | The Pleiades | *Mogari*, *Mongari*, *Kutpari*, *Thengari*, *Mundari.* | In one village (Karambi) it was pointed out at night. |
| 21 | | *Tiva*, *Purad*, *Hola*, *Purad* and *Mes* | These are stars west of Sirius in Canis Minor. This implement is used to drop the husk and seeds in the wind so that the husk flies away and the seeds are collected at the bottom of the implement. The myth of *Purad* (a bird) with *Mes* (two eggs) was recorded in only one village. Orion throws a stone which will hit the bird, so that the hunter can steal the eggs. Note that *Pudar* is to the east of Orion so the stone must follow a curved trajectory. |
| 22 | Scorpius | *Michu* | *Michu* is responsible for producing the dead body, *Murda*, mentioned in 23 below. |
| 23 | Leo | *Murda* (the dead body),<br>*Duf* (the drum bearer),<br>*Shika* (the procession leader),<br>*Ladavya* (the procession of crying women) | The body of Leo is considered the body of a dead person with the head located at Eta Leonis and one hand indicated by Algieba. The other hand is Regulus. The legs are formed by Delta Leonis and Theta Leonis.<br>Delta Leonis (*Asellus Australis*) is the pall bearer. Stars in Virgo form the funeral procession. The procession moves from west to east.<br>The whole procession of death is found in the sky. |
| 24 | The tail of the Scorpion | *Khala* | This constellation has a bright star in the centre of a circular pattern with faint stars surrounding it. It represents the animal-powered large grinding circles used in villages. |
| 25 | | *Irukmara*, *Irukna Mara* | This star is seen at 3 am, indicating the season to pick *Mahua* (*Madhuca longifolia)* flowers (i.e. March-April). |
| 26 | Centaurus | *Khayan,* | |

Table 6: Concepts encountered in Gond villages listed according to their frequency of occurrence.

| | Village (see pages 34-38). | | | | | | | | | | | | | | | |
|---|---|---|---|---|---|---|---|---|---|---|---|---|---|---|---|---|
| Serial No. | 1 | 2 | 3 | 4 | 5 | 6 | 7 | 8 | 9 | 10 | 11 | 12 | 13 | 14 | 15 | Total |
| Burial practices | E W | S N | S N | E W | S N | S N | S N | E W | E W | S N | S N | S N | S N | S N | S N | 15 |
| Referances to stars | Y | Y | Y | Y | Y | Y | Y | Y | Y | Y | Y | Y | Y | Y | Y | 15 |
| Orion and *Tipan* | Y | Y | Y | Y | Y | Y | Y | Y | Y | Y | Y | Y | Y | Y | | 14 |
| The *Saptarshi* Story | | Y | Y | Y | Y | Y | Y | Y | Y | Y | Y | Y | Y | Y | Y | 14 |
| Venus | | Y | Y | Y | Y | | Y | Y | Y | Y | Y | Y | Y | | Y | 12 |
| Comets | | Y | Y | Y | Y | Y | Y | | Y | Y | | Y | Y | Y | | 11 |
| The Milky Way | | | Y | Y | Y | Y | Y | Y | Y | Y | | Y | | Y | Y | 11 |
| Names of the Sun and Moon | Y | Y | | Y | Y | Y | | Y | Y | Y | | | Y | Y | Y | 11 |
| The Pleaides | | | Y | Y | Y | Y | Y | Y | | Y | Y | | Y | Y | Y | 11 |
| Shooting stars | | | Y | Y | Y | Y | Y | | Y | | Y | Y | Y | Y | Y | 11 |
| Duration of the month, and Leap years | | | | | Y | | | | | | Y | Y | Y | | Y | 5 |
| *Purad* or *Hola* (Canis Major) | | | Y | Y | | Y | | | | | | Y | | | Y | 5 |
| A glow around the Moon | | | | | | | | | | Y | | | Y | Y | Y | 4 |
| Names of the months | | | | Y | Y | Y | | | | | | | | | | 3 |
| *Samdur* | | Y | Y | Y | | | | | | | | | | | | 3 |
| Scorpius | | | | Y | | | Y | | | | | | Y | | | 3 |
| *Topli* | | | | Y | | | | Y | | Y | | | | | | 3 |
| The Pole Star | | | | | | Y | | | | Y | | | Y | | | 3 |
| *Bohari* | | | | | | | | | Y | | Y | | | | | 2 |
| *Medi* | | | Y | Y | | | | | | | | | | | | 2 |
| Names of cardinal directions | | Y | Y | | | | | | | | | | | | | 2 |
| *Irukna Mara* | | | | | | Y | | | | | | | | | | 1 |
| *Khala* | | | | Y | | | | | | | | | | | | 1 |
| *Khayan* (Centaurus) | | | | | | | | | | Y | | | | | | 1 |

| | Village (see pages 34-38). | | | | | | | | | | | | | | | |
|---|---|---|---|---|---|---|---|---|---|---|---|---|---|---|---|---|
| Serial No. | 1 | 2 | 3 | 4 | 5 | 6 | 7 | 8 | 9 | 10 | 11 | 12 | 13 | 14 | 15 | Total |
| *Koropadera* | | | Y | | | | | | | | | | | | | 1 |
| *Mrug* | | Y | | | | | | | | | | | | | | 1 |
| *Murda* and its companions | | | | Y | | | | | | | | | | | | 1 |
| *Nangel* | | | | | | | | | Y | | | | | | | 1 |
| Rainbow | | | Y | | | | | | | | | | | | | 1 |

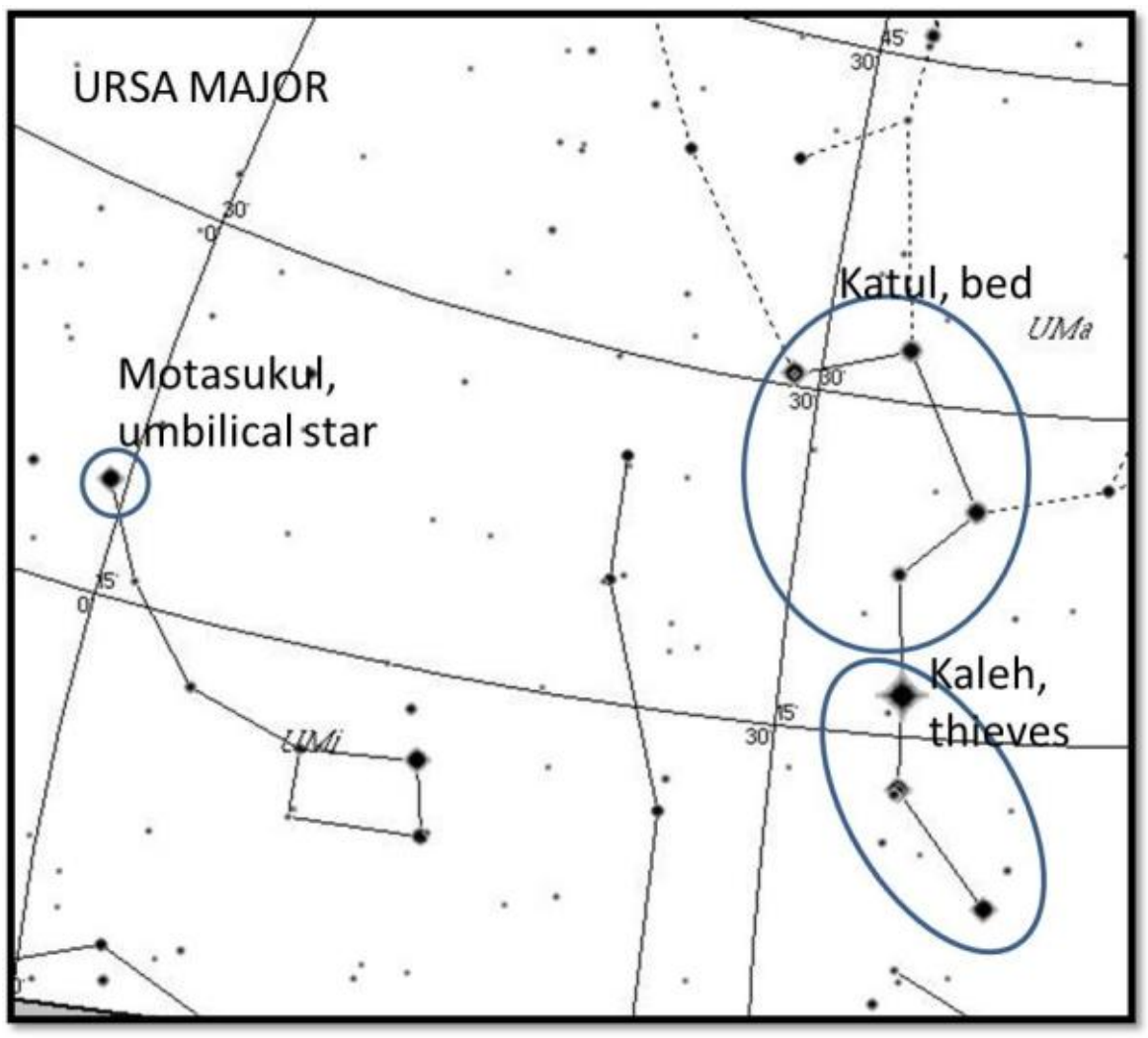

Figure 4: The region of Ursa Major.

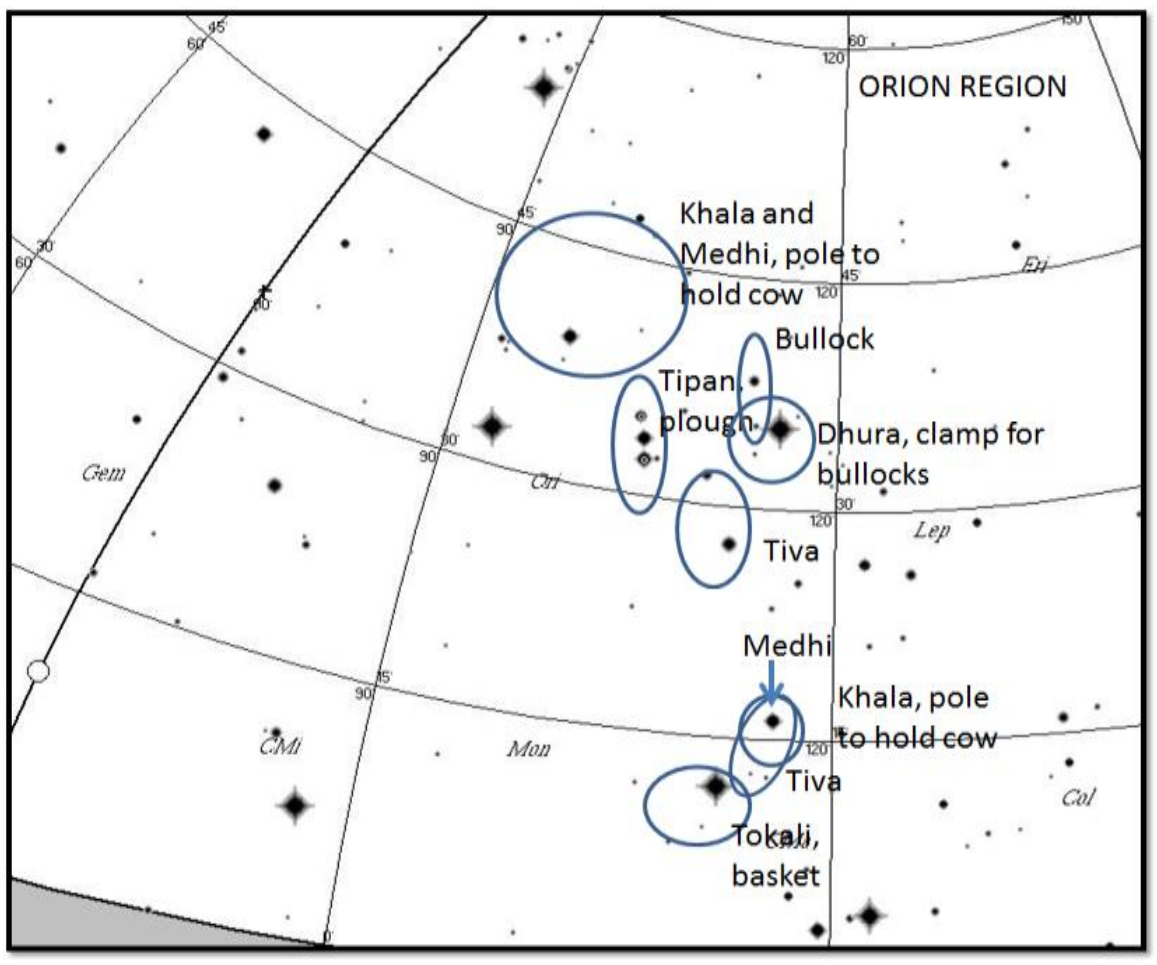

Figure 5: The region of Orion.

In Figures 4 to 9, we present maps of the sky with the Gond constellations marked on them.

## 5 CONCLUSIONS

The Gond community is clearly an ancient civilisation in its own right dating to a period well before the arrival of Dravidians in south India. Their genetics, lifestyle and mythologies all confirm this. In the present study we have analysed their astronomical beliefs and knowledge. Even our brief survey confirms that their astronomical beliefs were not influenced by later developments that occurred in India, and are sustained by various ancient ideas.

The stories and other astronomical information we collected can be divided into the following categories:

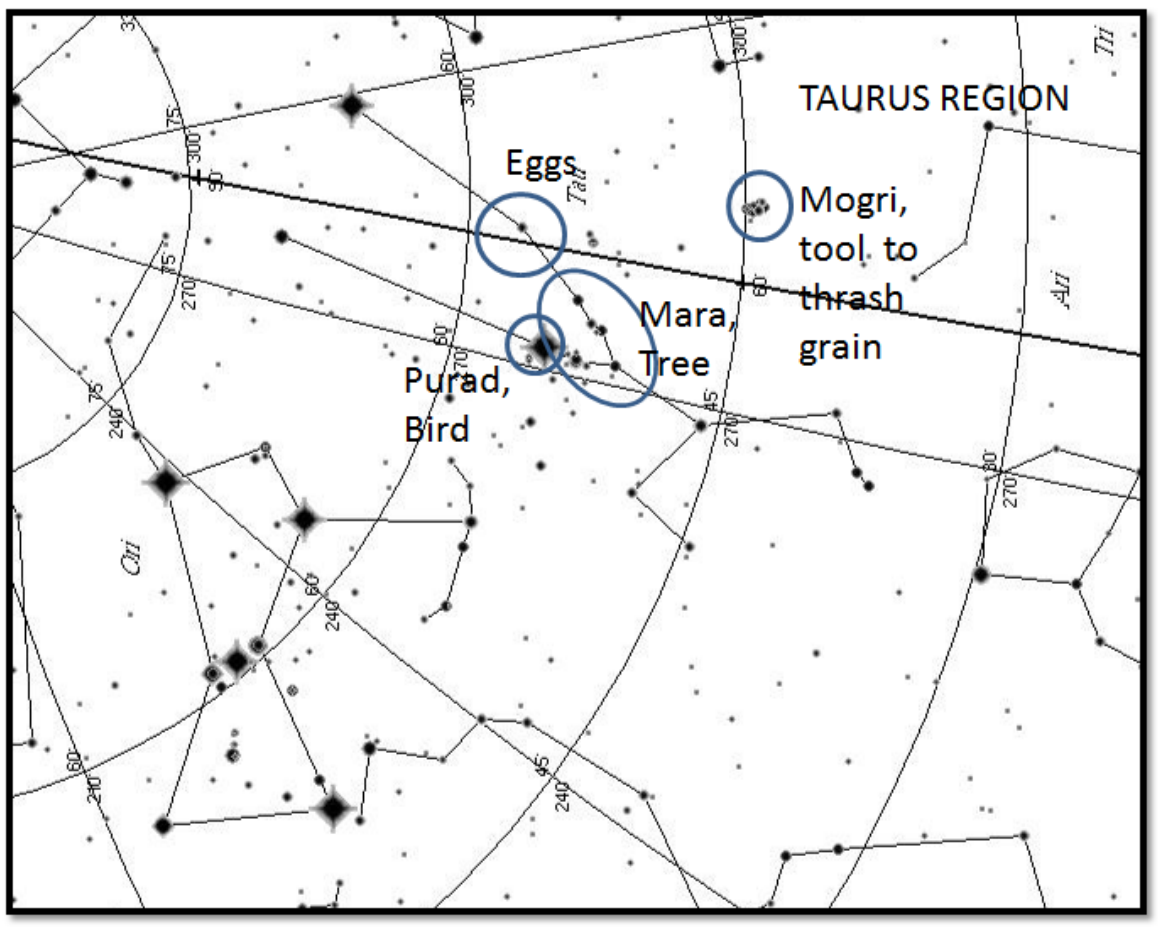

Figure 6: The region of Taurus.

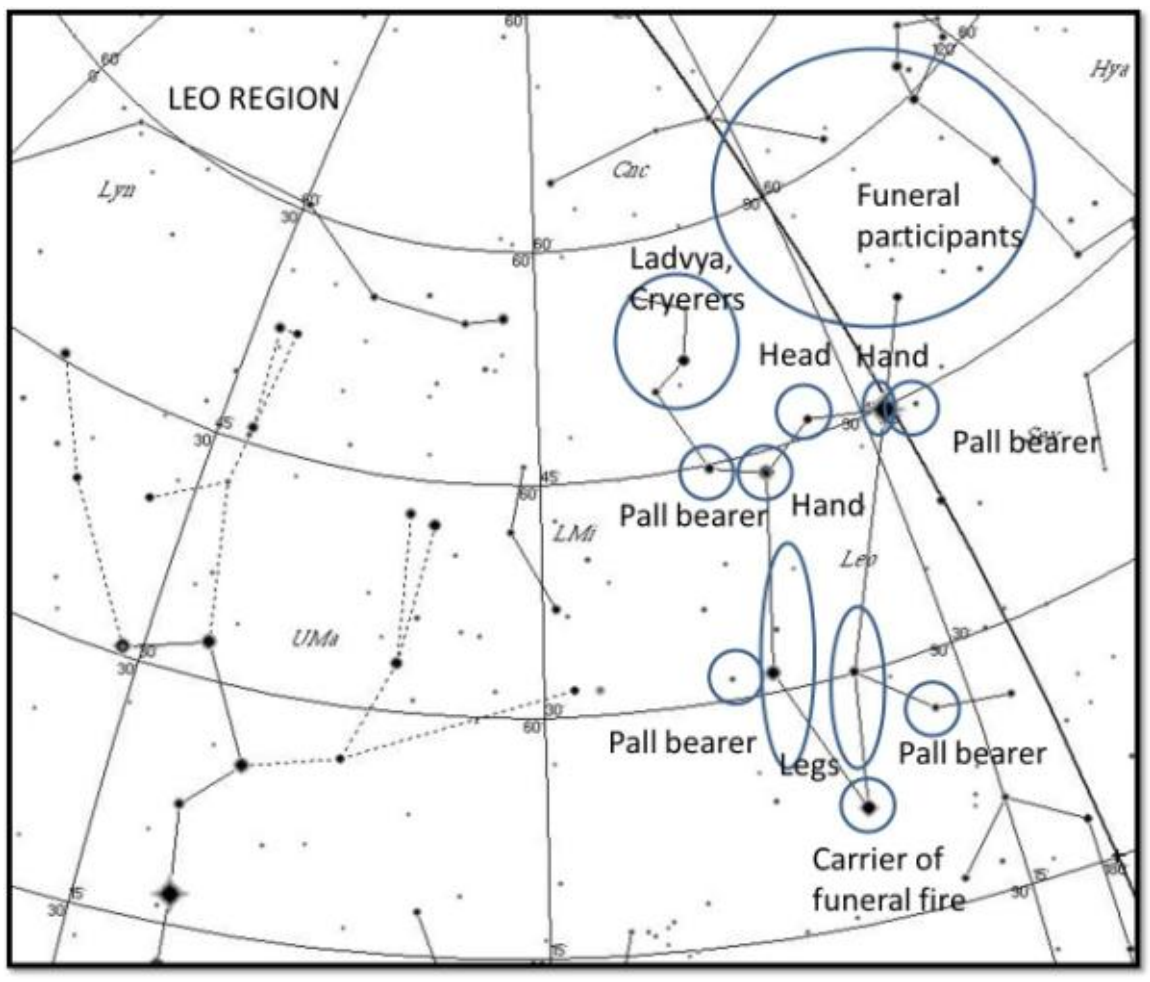

Figure 7: The region of Leo.

- Daily time-keepers – the Sun, the Moon, *Jevan Tara, Pahat Sukum,* the glow around the Moon
- Calendrical – constellation rise times, seasons
- Expression of human activities – *Tipan* and related star groups, *Murda*
- Mythological – comets, the Milky Way, shooting stars
- Cosmogonical – *Saptarshi*

It is clear that the Gond people used astronomy for a variety of purposes from simple daily and annual time-keeping to projecting their life in the skies and cosmogony. Note that all their festivals are based on the lunar calendar (Table 1). However, they do not seem to have used it for navigation. We also did not find a single instance where they numbered their calendars beyond the three years needed to add the intercalary month. Clearly, given the scope of Gond culture, this must be more of a memory loss rather than a tradition. Together, therefore, even this sample study indicates Gond interest and perception of astronomy with underlying mythologies and practical ideas. That they had a very well-pursued idea of the intercalary month suggests sensitivity to the seasons and synchronisation of solar and lunar calendars. Their firm and commonly-held belief and knowledge of comets which typically appear only a few times in a century also indicates a continuing tradition of astronomical and other observations. However, they clearly lacked any knowledge of eclipses, which are a relatively common (and periodic) phenomenon, indicating either an absence of very keen observations, or equally likely, a conscious decision to ignore information that did not easily fit their world view.

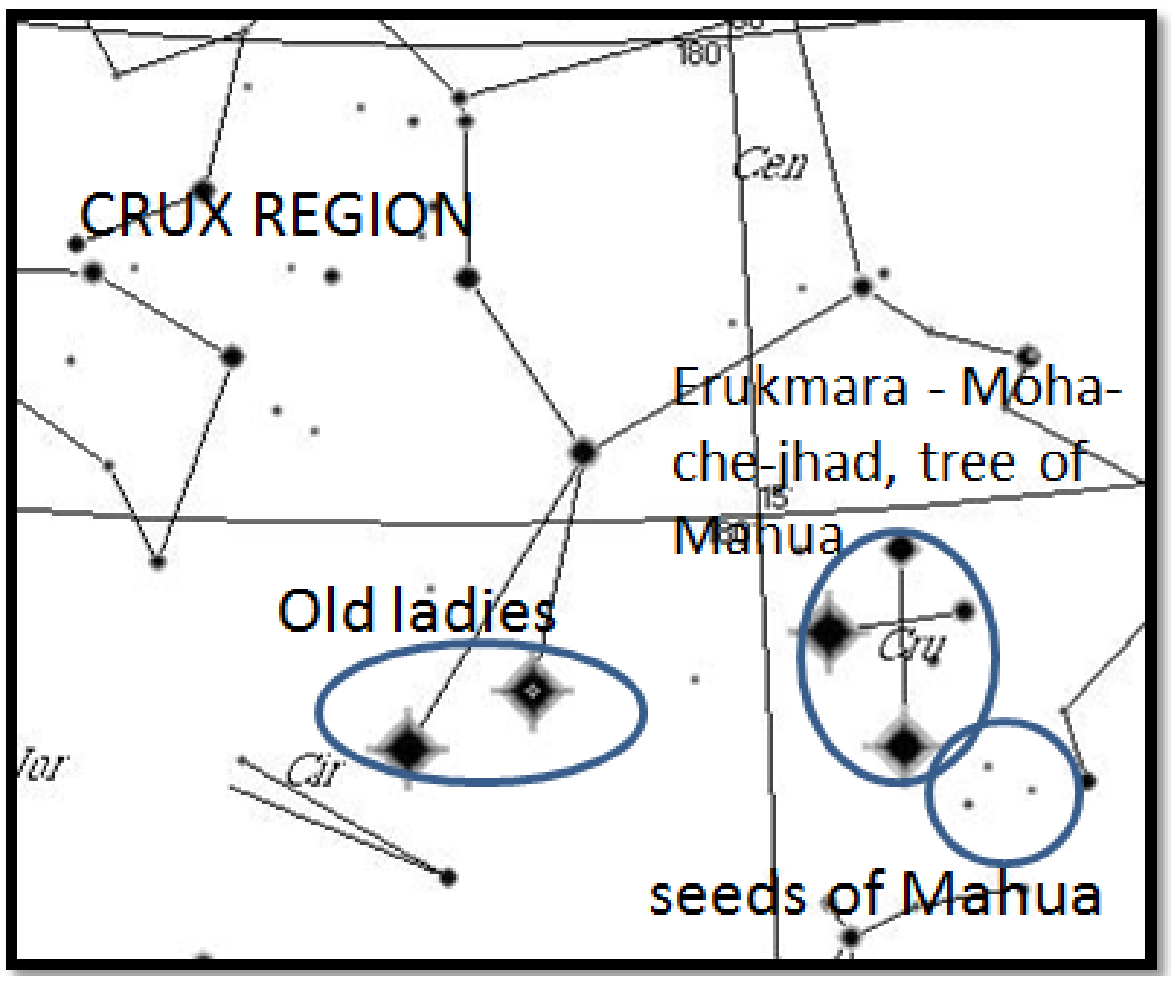

Figure 8: The region of Crux.

Another marked feature of their astronomy is the absence of gods or super-humans, except for *Saptarshi*. This again suggests that in spite of having an agrarian lifestyle, the Gonds were not given to grandiose speculations about the heavens and events occurring in the sky.

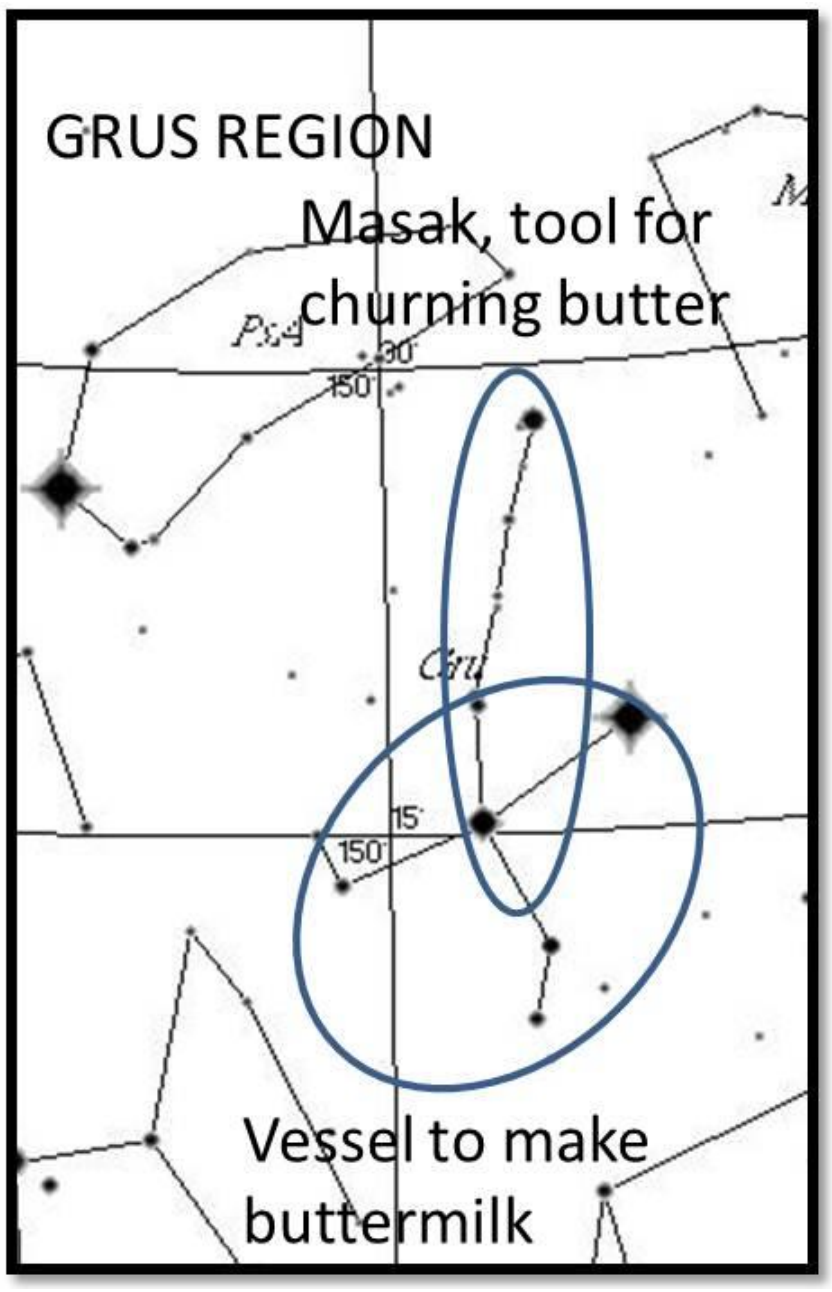

Figure 9: The region of Grus.

Another aspect of the observations is the Gonds' lack of interest in constellations such as Cassiopeia, Aquila, Gemini, Bootes, Cygnus and Sagittarius. These contain bright stars yet the villagers could not even identify them in the planetarium. It is significant that apart from Sagittarius, all of these constellations lie in the northern sky, north of the northern-most point of sunrise (Mahendra Wagh, private communication), which suggests a lack of interest in northern constellations where the Sun does not travel.

The naming of Polaris as the umbilical star suggests cosmogonical ideas based on the Pole Star as the centre of the Universe and humans. Such an interpretation of the heavens is also found in the *Surya Siddhanta*.

One more interesting feature of Gond astronomy is that their observations extended all the way down to Crux and Grus, confirming that the Gonds were keen observers of their own local sky and did not import astronomical ideas from people living farther south.

As regards planets, they seem to have noticed only Venus and identified it as both the Morning and the Evening Star.

All in all therefore, it seems that Gond astronomy had its roots in early farming needs and was designed several thousand years ago when Polaris was not yet the Pole Star and *Saptarshi* was circumpolar, which happened around 1000 BC. This reinforces the general consensus that the origin of the Gonds is much older than previously thought. There also seems to have been little later modification of this basically utilitarian approach to life and environment which is a hallmark of Gond traditions.

It would be useful to follow up this study in greater detail, and also endeavour to compare the astronomical views of the Gonds with those of other Indian tribes.

## 6 NOTES

1. In Hindu society, the term *Gotra* means clan. It broadly refers to people who are descendants in an unbroken male line from a common male ancestor (after Wikipedia).
2. In Madhya Pradesh the Gond calendar is designed by Chaitanya Kumar Sinha of Rajanandgaon. In Maharashtra it is published by Tiru Moreshwar Tukaramji Kumare and Tiru Sampatji Kannake Ballarshah and is printed by Ohmkar Graphics in Chandrapur.
3. This is a deep N, produced by using the soft (back) of the palate rather than the 'normal' N that is produced by using the hard (front) of the palate.

4. A short film about this April 2012 visit of the Gonds to the planetarium at the Raman Science Centre in Nagpur can be viewed at: www.tifr.res.in/~archaeo

## 7 ACKNOWLEDGEMENTS

The authors wish to acknowledge the support of the Jamsetji Tata Trust in carrying out this work. We are particularly grateful to our friend Mr Kishore Menon who has converted our confused writing into a very readable manuscript and who has been an enthusiastic friend through this and many other journeys. Without his hard work, this paper would have been significantly more difficult to read. We remain grateful to him. We are also equally grateful to Professor Wayne Orchiston who worked untiringly on the manuscript ensuring that every sentence in the paper and the tables were properly correlated and there were no errors. This alone ensured that the paper is comprehensive and has added immensely to the consistency and quality of the paper. We thank him for all his effort and commitment. We also wish to thank Professor Sir Arnold Wolfendale, FRS, for his helpful comments and advise. We also wish to thank our sincere and committed driver, Vilas Khoje, who drove us around with dedication and also helped us in other tasks. We also wish to acknowledge our many helpful discussions with Ms Nisha Yadav. Mr Phatak and other staff members of the Raman Science Centre were a great help in our programme at the planetarium, and we are grateful to them for their assistance. We also wish to thank Ms Harini Calamur for making the film during the visit of the Gonds to the planetarium.

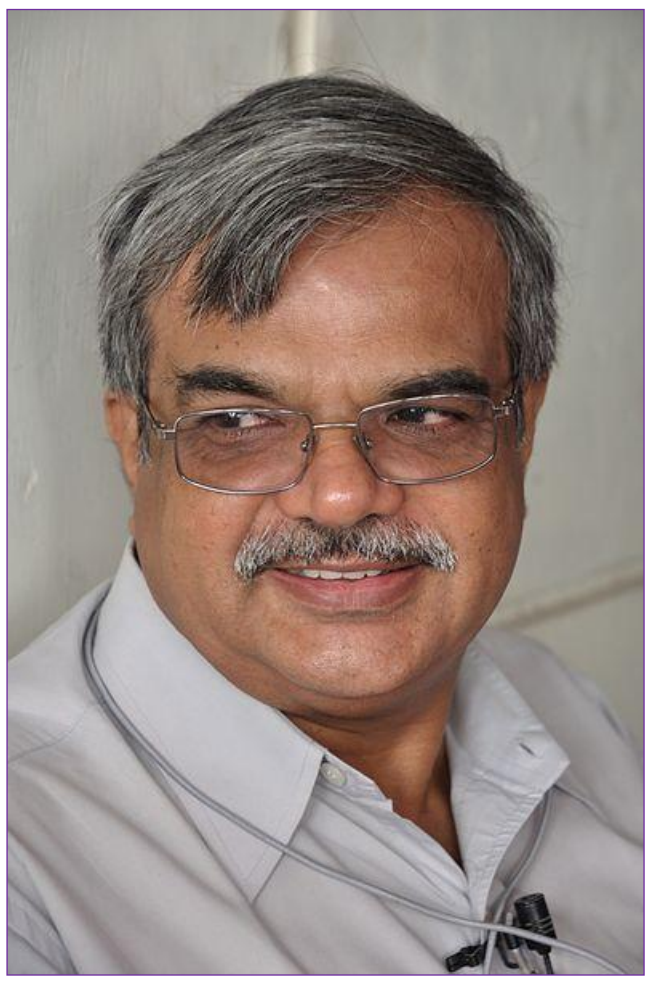

Mayank Vahia is a Professor in the Department of Astronomy and Astrophysics at the Tata Institute of Fundamental Research in Mumbai. He began his career making satellite-based astronomical instruments for high energy astrophysics. He was the Director of the Nehru Planetarium in Mumbai for a year in 2000-2001. He is currently the National Coordinator of the Astronomy Olympiad in India. In recent years he has become interested in the origin and growth of astronomy in the Indian subcontinent. He has been working on prehistoric records of astronomy in India and is deeply interested in documenting astronomical beliefs amongst Indian tribes. Mayank has published more 200 papers in research journals, about 35 of which are in history of astronomy.

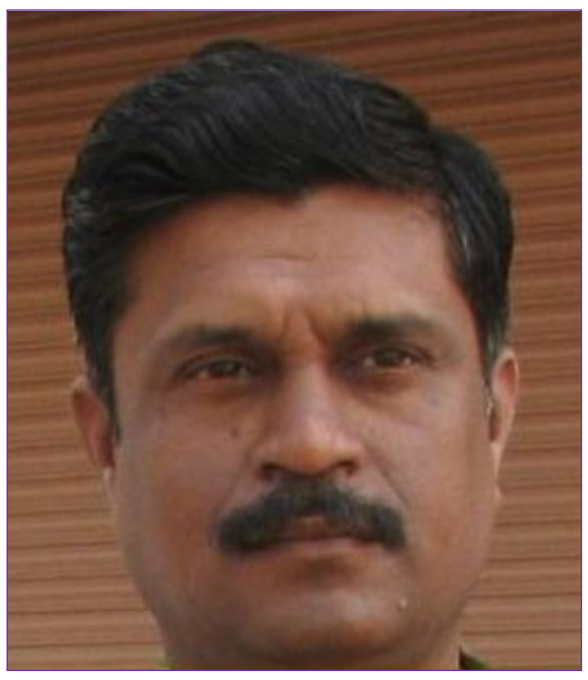

Ganesh Halkare is an advocate in Amravati, a town in Maharashtra. He is also a post-graduate degree holder in Archaeology and Anthropology from Nagpur University. He has a deep interest in tribal education, particularly in the removal of superstition among the tribe members. He is also deeply interested in tribal anthropology and is highly respected amongst the tribesmen for his work in ensuring that they are aware of and can exercise their rights within the nation state. He also has been a regular columnist on constitutional rights. Ganesh has published more than a dozen research papers on the archaeology of the Nagpur region in Indian journals and conference proceedings. He is now working on the astronomy of other tribes in the Nagpur region.